# Furthering Asteroid Resource Utilization in the Next Decade through Technology Leadership

A Planetary Science and Astrobiology Decadal Survey 2023-2032 White Paper

**Lead Author:** Chris Lewicki (chris@Interplanetary.Enterprises) (Interplanetary Enterprises, LLC)

**Co-authors:** Amara Graps (Baltics in Space and Planetary Science Institute), Martin Elvis (Center for Astrophysics | Harvard & Smithsonian), Philip Metzger (University of Central Florida), Andrew Rivkin (Applied Physics Laboratory, Johns Hopkins University)

**Abstract:**

A significant opportunity for synergy between pure research and asteroid resource research exists. We provide an overview of the state of the art in asteroid resource utilization, and highlight where we can accelerate the closing of knowledge gaps, leading to the utilization of asteroid resources for growing economic productivity in space.

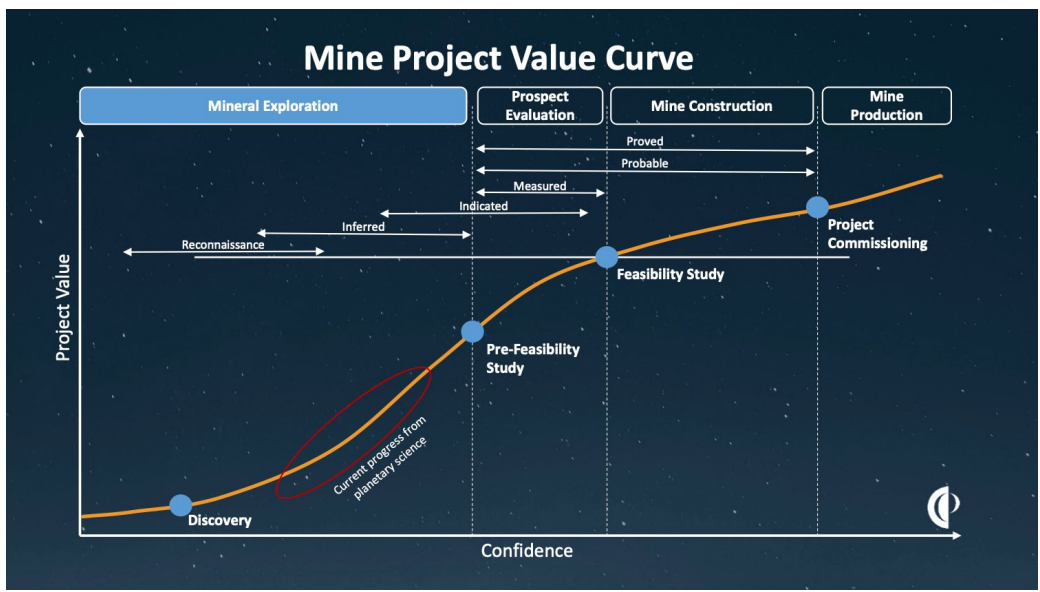

*Mine Project Value Curve - Project Value as a function of mineral resource confidence, and associated terrestrial industry terms.*

**Executive Summary**

The study of asteroids as a way to understand Solar System origins and dynamic evolution is accelerating with recent science missions including OSIRIS-REx, Hayabusa 2 and soon to be launched missions including Psyche, Lucy, DART, MMX and Janus. Simultaneously there has been a rapid growth of activity around *using* the vast resources of the asteroids. Many of the science mission results also contribute to understanding these objects as potential orebodies. A significant opportunity for synergy between pure research and asteroid resource research has opened up, with pure and applied research each benefiting from the other. This white paper provides a brief overview of the state of the art in asteroid resource utilization, and highlights where further research can help accelerate the closing of science knowledge gaps, leading to the utilization of asteroid resources for growing economic productivity in space.

**Asteroid Resources**

Asteroid Resources are vast. The iron in the Main Belt has some 10 million times the known resources on Earth[1]. Water and other volatiles, as well as precious metals, are also abundant among the asteroids. For a long time, though, the prospect of actually harnessing these resources has seemed remote. Now "Mining the Sky" is no longer seen as a fantasy, but as a realistic, though challenging, enterprise. Starting about a decade ago, there have been a number of well-publicized start-ups with asteroid mining as their goal: Planetary Resources, Deep Space Industries, TransAstra and others. These commercial pursuits have been supplemented by a space-resource focused degree program at the Colorado School of Mines, and graduate seminars at the University of Central Florida, among others.

The near-term hurdles are mostly financial. Development of the necessary technology needs too long a horizon (>5 years) for most investors. The US government has often brought down risk for entrepreneurs by investing in new technology[2]. NASA is already supporting asteroid resource activities via the SBIR and NIAC programs[3]. Keeping in mind that NASA is charged to "seek and encourage, to the maximum extent possible, the fullest commercial use of space"[4], and its strategic goal #3: "Address national challenges and catalyze economic growth"[5], support of asteroid mining perfectly fits NASA's mandate. More directly, Planetary Science will directly benefit from asteroid resource utilization. The same technologies developed for asteroid resource use will also bring great benefits to planetary science by enabling many more deep space missions within the same cost envelope. Similarly, planetary science will inform the prospecting and mining of these resources, enabling more efficient prospecting.

---

[1] "Mining the Sky: Untold Riches from the Asteroids, Comets, and Planets", John S. Lewis (Reading, MA: Addison-Wesley, 1996); Elvis, M., and Milligan, T., "How much of the solar system should we leave as wilderness?" Acta Astronautica, (2019), 162, 574.
[2] see e.g. "The Entrepreneurial State", Mariana Mazzucato (2013, Anthem Press.)
[3] E.g. Demonstration of "Optical Mining" For Excavation of Asteroids and Production of Mission Consumables: https://www.sbir.gov/sbirsearch/detail/887233
[4] Title 51, U.S. Code § 20112. Functions of the Administration: https://www.law.cornell.edu/uscode/text/51/20112
[5] NASA Strategic Plan 2018: https://go.nasa.gov/2ZBjwbV

Here we flow down from the goals, through the currently recognized science knowledge gaps, to generate a list of activities that NASA, NSF, USGS and other agencies could undertake in the 2023-2032 decade to further asteroid resource utilization[6].

**II. An International Asteroid Mining Community Roadmap to address Asteroid Resource Science Knowledge Gaps.**

The decade of the 2010s saw a global shift of interest towards private space industry endeavors. Due to their different funding streams (private versus public), communication and interaction between private space workers and planetary scientists was relatively limited until the middle of the decade. From the point of view of the asteroid scientists, why weren't the companies hiring asteroid specialists?[7] From the point of view of the companies, their concern was financial: how to develop the research and technology for mining in the far future with limited R&D support, while financially supporting their company today? Asteroid scientists are an expensive resource too. Academic research to further asteroid mining has been extremely limited.

To address this problem, A. Graps convened the international planetary science community with asteroid mining companies and their funders in the first workshop on "Asteroid Science Intersections with in-space Mine Engineering" (ASIME) in September 2016 in Luxembourg[8]. The workshop themes included astronomical surveys, near-surface and on-surface asteroid operations. The questions by the companies were collected by J.L. Galache and the answers were provided by the asteroid researchers, which yielded 'science knowledge gaps' (SKGs)[9]. The ASIME-derived SKGs provided the first *community-developed* scientific roadmap supporting asteroid mining. The theme of ASIME 2018 was asteroid composition and related SKGs with the community's focus on the OSIRIS-REx and Hayabusa 2 asteroid missions[10]. The next refinement of the SKGs is planned in the third ASIME workshop (delayed until April 2021).

---

[6] The ethical challenges arising from the use of asteroid resources are addressed in a companion white paper (Asteroid Resource Utilization: Ethical Concerns and Progress, Rivkin et al. 2020).
[7] "Why Commercial Space Startups Need Scientists", Elizabeth Frank, https://bit.ly/33xR2ke
[8] *Asteroid Science Intersections with In-Space Mine-Engineering (ASIME) 2016.* September 21-22. Luxembourg City, Luxembourg. http://fmispace.fmi.fi/index.php?id=asime16
[9] Graps, A. + 30 Co-Authors. 80-page White Paper ASIME 2016 outcome with Science Knowledge Gaps: "Answers to Questions from the Asteroid Miners" (https://arxiv.org/abs/1612.00709)
[10] Graps, A. +43 Co-Authors. 65-page White Paper ASIME 2018 outcome: "Answers to Asteroid Composition Questions from the Asteroid Miners" (https://arxiv.org/abs/1904.11831)

**III. State of the Art in 2020 – Asteroid Resource Science Knowledge Gaps**

The ASIME process has identified the many steps in the resource utilization process. While progress has been rapid[11,12], many require further development[13]. The six "Science Knowledge Gaps" identified by ASIME 2016 map to some fundamental questions of asteroid science: (1) better mapping of meteorite classification to asteroids, (2) dedicated near-Earth asteroid (NEA) discovery and follow-up, (3) understanding of granular material dynamics and low gravity, (4) identifying the most accessible objects, (5) making more certain links between predicted NEA source regions and actual physical properties, and (6) adequate data on the particle size-frequency distribution of asteroid regolith and deeper interior.

Here we provide a first list of 2020 Decadal science knowledge gaps for asteroid resources, one that can be refined by a more formal study.

1. *Knowledge of the Asteroid Population.* For the first asteroid resource ventures it is the near-Earth objects (NEOs) that are relevant, as they are on average more accessible than the Main Belt and other populations. Only a few percent of NEOs are anticipated to be rich in the hydrated minerals for the extraction of water,[14] the favored first product of asteroid mining. Hence expanding knowledge of the NEO population is a prerequisite for prospecting asteroid resources. The concomitant gain for small body studies in general is clear.

2. *Prospecting Resource Richness.* Terrestrially, benefitting from the more extensively developed (and comparably data-rich) field of geology, the mineral and energy industries have developed standards for demonstrable knowledge of the availability of resources of interest. Mineral Resources are subdivided, in order of increasing geological confidence, into *Inferred, Indicated*, and *Measured* categories. Thus far, virtually all resources beyond Earth's surface lie are merely Inferred or Indicated, which are insufficient to make useful economic viability estimates. In mining terms, a "Feasibility Study" progresses through increasingly informed analyses from "Order of Magnitude", graduating to "Preliminary Feasibility", before potentially achieving a "Detailed Feasibility Study" which is used to inform funding decisions[15]. Asteroid resources are mostly in the Order of Magnitude stage. Later stages represent more substantiated resource confidence (see cover figure). Applying these concepts to space resources leads to the need for bulk characterization, homogeneity and to being able to cross-reference inferences from related asteroid populations, whether those measurements are made directly or indirectly from meteoritics, astronomical spectroscopy, sample return, or in-situ measurements. Being able to infer the content of asteroids at the Preliminary Feasibility level would benefit studies of scientifically interesting asteroids too.

---

[11] Graps, A. Ten Asteroid Mining Milestones 2018-2019 https://bit.ly/33w35yr, 2019-2020 https://bit.ly/2RqYkAP
[12] Aerospace America, Dec 2019 "International momentum for space resources ramps up" https://bit.ly/3km3tqk
[13] Galache, J.L and Graps, A.L. "What do you need to know to mine an asteroid?" https://bit.ly/2ZEzWQS
[14] Rivkin A., DeMeo F. JGR Planets (2018) "How Many Hydrated NEOs Are There?" https://bit.ly/33vDnKL
[15] Mining Feasibility Study https://en.wikipedia.org/wiki/Mining_feasibility_study accessed Sept 2020

3. *Deep Space Prospecting/Exploration*. The need to find a few good asteroids means using in situ measurements of tens of candidates. So many missions must then be cheap, and so small and largely autonomous, in order to be economic. NASA's recent inclusion of small deep space missions and their supporting technology will provide commensurate benefit here.
4. *Handling an asteroid.* Operating spacecraft near an irregular, possibly tumbling asteroid poses problems. Hayabusa 1, 2, Rosetta and OSIRIS-REx have begun to retire this risk. Attaching to the surface of an asteroid, rather than touch-and-go or tumbling landings, has not yet been attempted, but is necessary for large scale asteroid mining. The ability to land and safely access the asteroid sub-surface would clearly benefit planetary science.
5. *In situ extraction.* Many asteroids are "rubble piles" complicating excavation methods, and making it difficult to gain a strong purchase on the surface. The evolution of small solar system bodies may well be influenced by their granular physics. Yet the physics of granular bodies in zero-g is little explored and lacks a predictive theory.
6. *Beneficiation.* Concentrating the desired resource from the raw material (refining) will require both adapting Earth-based methods to 0-g, and developing new methods which leverage the unique environment (e.g. TransAstra Optical Mining™).

**IV. Near-term Activities to Further Asteroid Utilization**

Currently global space programs visit about one small body every two years. The smallsat revolution applied to asteroid prospecting could enable in situ science missions to NEOs by the dozen in the 2030s. The increase in the available volume and variety of data could be profound. With the development of a sustained space resource industry, it is anticipated that private funding for asteroid science studies will outweigh public sources[16]. Historically, much of (if not most) Earth science data has come from economic geology rather than purely academic research campaigns[17]. With up to 95% of geoscientists supported directly or indirectly by economic geology[18], economic planetary science could likewise improve the health of the system. Economic development in the region called "space" has been found to be correlated to scientific productivity in space[19]. These considerations indicate that resource utilization missions will support science in more ways than just the datasets they collect, producing great returns in planetary science. Addressing the 2020 Decadal science knowledge gaps can be addressed in the 2023-2032 timeframe with focus in the following areas:

---

[16] Metzger, P. T. (2016). "Space development and space science together, an historic opportunity." Space Policy, 37, 77-91.

[17] Metzger, P. T. "Economic planetary science in the 21st century." Planetary Science Vision 2050 Workshop, *LPI* Abstract #1989 (2017): 8126.

[18] U.S. Bureau of Labor Statistics website. https://www.dol.gov/

[19] Gantman E.R. (2012), *Scientometr.* 93, 967–985.

[1] **Enhanced access to asteroid data.** Collating all existing and future data on asteroids, especially NEOs, into easily accessible public databases could speed up resource assessment for mining, while also providing a valuable tool for planetary science. The international asteroid physical properties databases provide macro physical properties[20,21,22]. However, further details on the individual asteroid's surface properties are needed and can be addressed via a multiwavelength collation such as that in the Asteroid Regolith Database[23] and in asteroid regolith simulants[24]. Interoperability of these databases to allow queries to be defined with parameters in any combination of databases, including using APIs would be helpful.

[2] **Our knowledge of the population of NEOs** will be essentially completed down to ~50 meters diameter by 2032 in terms of discovery and orbit determination. In the mid-IR the NASA NEO Surveillance Mission (NEOSM) and, in the optical, with the NSF Vera Rubin Observatory (VRO) will accomplish this task. NEOSM is particularly useful for asteroid resources as it is unbiased against dark objects and can provide good size estimates from thermal modeling, and so improved resource estimates. In combination VRO and NEOSM give the NEO albedos. Carbonaceous, and potentially water-rich, asteroids usually have low albedo, separating them from the bulk of the asteroid population.

Finer-grained classification of NEOs could come from greatly increased ground-based spectroscopy to rapidly *classify their types by the thousand*. Such data would allow prospecting efforts to focus on the minority of promising NEOs, and would advance asteroid science. NASA and NSF investments in existing 8-meter class telescopes for both asteroid science and resource reconnaissance could jointly provide a new mission for the present generation of these telescopes will be outclassed by the new generation of 30-meter-class telescopes around 2030. More detailed *infrared spectroscopy of NEOs* could provide many more diagnostics, notably the 3 micron water feature. While IR spectroscopy of faint asteroids is very challenging from the ground, a spectrometer on a meter-class space telescope may provide direct knowledge of surface composition, including water content, for many more NEOs. Study of such a mission for launch at the end of the 2023-2032 decade is desirable.

[3] **Determining asteroid-appropriate economic resource estimates**. Several NEOs have signatures of hydrated minerals on their surfaces, and so may be the first targets of asteroid mining efforts. However, is an asteroid's surface representative of its interior? This matters for resource assessment, but also for understanding asteroid evolution. Can remote sensing provide diagnostic information about its interior? Measurements from OSIRIS-REx and Hayabusa 2 can be used to address this question, both from in situ data and from study of the returned samples. Hayabusa 2's samples will be of particular interest as they contain material

---

[20] Asteroid Physical Properties archives @ JPL Small-Body Database https://ssd.jpl.nasa.gov/sbdb_query.cgi
[21] Asteroid Physical Properties Database @ Lowell Observatory https://asteroid.lowell.edu/astinfo/
[22] MP3C @ OCA (France) https://mp3c.oca.eu/catalogue/index.htm
[23] Graps, A. L., Vaivode, M., Consolmagno SJ, G. J., and Britt, D. T.: An Asteroid Regolith Database for ISRU, Europlanet Science Congress 2020, online, 21 September–9 Oct 2020, EPSC2020-939, 2020
[24] Exolith Lab : https://sciences.ucf.edu/class/exolithlab/

excavated from the sub-surface. Data from many additional NEOs will be needed to generalize these results.

[4] **In-situ spacecraft exposure experiments to mitigate dust risks.** The deleterious effects of asteroid dust — "fines" — on thermal, radiating and optical surfaces and mechanisms are likely to be significant for mining and science missions alike. Understanding these effects will allow mitigations to be implemented to create mission resilience. Regolith dust particles are charged by the solar wind plasma and lofted by electrostatic forces, altering the regolith surfaces of asteroids with consequences for both scientific and mining missions. Our understanding of dust lofting has recently been revolutionized after 40 years[25]. The Patched Charge Model[26], shifts our view from the regolith surface to charging *within microstructures* inside the regolith. Large, negative charges can lead to explosive ejection of dust. In situ tests of this model are needed for basic physics understanding and for developing mitigation strategies for asteroid surface operations.

[5] **In-situ extraction experiments.** Micro-g experiments on simulants, mini-moons, on recently-collected samples, or in situ in the asteroid environment, allow for exploration of techniques for the extraction of useful materials, and scientifically interesting minerals, from raw asteroid material. A myriad of challenges, and likely unpleasant surprises, await in beneficiating material in this unfamiliar environment. At industrial scales the characterization of waste and the concentration of natively toxic products may be of concern for autonomously sustainable and environmentally-conscious processes. Initially these investigations may be hosted on board a kinetic penetrator, such as the tunable diode laser spectrometer and resistive calorimeter proposed in Planetary Resources' inaugural Exploration Mission.[27,28] Follow-on investigations can refine, and scale-up successful experiments to larger, higher-throughput prototypes until commercially viable designs are proven and implemented.

[6] **Perturbation of "global" asteroid equilibrium**. Asteroid "geotechnical" properties are largely unknown, but critical for designing and operating infrastructure on their surfaces. Most or all NEOs are thought to be rubble piles, including many structural voids, and many have rotation periods and shapes that are near or at stability limits after aeons of Solar System processing. Mining operations, or even scientific drilling, could cause large-scale resurfacing and reshaping of NEOs, including mass loss, and possible loss of mission. In order to understand these effects, avoid consequences, and exploit opportunities, research and experiments to understand the effects of "global" perturbations to the equilibrium of a small body will be very productive. These investigations will likely include microgravity granular physics, or global effects of changes in angular momentum magnitude and/or vector[29]. Experimentation on

---

[25] See 2020 Decadal Survey White Paper: X. Wang et al: "Electrostatic Dust Transport Effects on Shaping the Surface Properties of the Moon and Airless Bodies across the Solar System"
[26] Wang, X., J. Schwan, H.-W. Hsu, E. Grün, and M. Horányi (2016), Geophys. Res. Lett., 43, 6103–6110.
[27] Arkyd-301, https://www.planetaryresources.com/missions/arkyd-301/ accessed Sept 2020
[28] "The Planetary Resources Exploration Plan", Unpublished internal document, 2017
[29] WRANGLER - Capture and De-Spin of Asteroids & Space Debris, https://go.nasa.gov/3kfV33E

simulants at small scale on the ISS, then on larger mini-moons and/or asteroids would address these issues.

[7] **Research and provision of asteroid resource simulants and analogs.** Meteorites provide many direct samples of carbonaceous asteroid material, but are too rare to allow extensive experimentation in asteroid extraction, purification and manufacturing activities. Fortunately, NASA has extended its funding for Martian, Lunar, and asteroid analogs and asteroid simulants [30,31]. Continued improvement and support of simulants to ensure their availability for scientific and commercial research would facilitate important experiments in the field.

[8] **Solar-orbiting debris created from asteroid mining**. The first NEOs to be mined will have very low delta-v to access and so will be in closely Earth-like orbits. Debris from mining could become an issue in Cislunar space[32]. An "Environmental Impact Study" on the evolution and transport of asteroid debris clouds could identify potential concerns induced by industrial-scale activity. The Double Asteroid Redirection Test (DART) Mission will provide a first test case[33]. Continued research, modeling, and experimentation will help to inform potential limits to future asteroid-located processes.

**V. Recommendations for the Next Decade**
**We recommend that support of asteroid science that helps to close these knowledge gaps continue. We recommend that NSF, NASA, USGS, and other agencies consider specific support for such research. We also recommend that interested parties revisit the 2016-era ASIME knowledge gaps to consider what has been retired in the wake of data from Bennu and Ryugu, and which remain or have been added.**

In particular, NASA, NSF, USGS and other agencies should:
1. Implement the Near-Earth Object Surveillance Mission (NEOSM) and the Vera Rubin Observatory (VRO) NEO surveys;
2. Expand ground-based NEO characterization, possibly using 8-meter-class telescopes;
3. Develop a near- to mid-IR spectroscopic in-space capability for NEO surface characterization;
4. Implement low-cost sample return missions;
5. Study dust charging on asteroid surfaces and mitigation techniques;
6. Initiate resource extraction experiments with simulants and in situ;
7. Perform experiments on granular materials and bodies in zero-g;
8. Support asteroid simulant production;
9. Study the potential environmental impact of asteroid mining.

---

[30] NASA SBIR "Task-Specific Asteroid Simulants for Ground Testing" Deep Space Industries https://bit.ly/3kkIvZ0
[31] Planetary Simulant Database: UCF/DSI CI Carbonaceous Chondrite Simulant, https://bit.ly/35EMPho
[32] Fladeland, L., Boley, A. C., & Byers, M. (2019). Meteoroid Stream Formation Due to the Extraction of Space Resources from Asteroids. arXiv preprint arXiv:1911.12840.
[33] On the delivery of DART-ejected material from asteroid (65803) Didymos to Earth, arXiv:1912.09496